
\input  harvmac


\def\d{{\rm d}}
\def\e{{\rm e}}
\def\civ{\lambda}
\def\tr{{\rm tr}\,}
\def\ud{\half}
\def\sigmt{\tilde\sigma}
\noblackbox

\centerline{\titlefont $O(N)$  Vector Field Theories in the Double}
\centerline{\titlefont Scaling Limit}
\vskip1.3truecm
\centerline{J. Zinn-Justin}
{\baselineskip14pt\centerline{Service de Physique Th\'eorique, CE-Saclay}
\centerline{F-91191 Gif-sur-Yvette Cedex, FRANCE}
\centerline{zinn@poseidon.saclay.cea.fr}}
\footnote{}{$^1$Laboratoire de la Direction des Sciences de la Mati\`ere du
Commissariat \`a l'Energie Atomique}
\footnote{}{$^2$Lecture given at the   \lq\lq XXV$^{\rm th}$ Symposium
on the Theory of Elementary Particles'' (Gosen 1991), SPhT/91-185}
\vskip.3in
{\it Abstract.} $O(N)$ invariant vector models  have
been shown to possess non-trivial scaling large $N$ limits, at least
perturbatively within the loop expansion, a
property they share with matrix models of 2D quantum gravity. In contrast
with matrix models, however, vector models can be solved in arbitrary
dimensions. We present here the analysis of field theory
vector models in $d$ dimensions and discuss the nature and form of the
critical behaviour. The double scaling limit corresponds for $d>1$ to a
situation where a bound state of the $N$-component fundamental vector field
$\phi$, associated with the $\phi^2$ composite operator, becomes massless,
while the field $\phi$ itself remains massive. The limiting model
can be described by an effective local interaction for the corresponding
$O(N)$ invariant  field. It has a physical interpretation as describing
the statistical properties of a class of branched polymers.\par
It is hoped that the $O(N)$ vector models, which can be investigated in their
most general form,  can serve as a test ground for new ideas about the
behaviour of 2D quantum gravity coupled with $d>1$ matter.
\nref\rAMV{Anderson A., Myers R.C. and
Perival V., Phys. Lett. {\bf B254}, 89 (1991); Nucl. Phys. {\bf B360}, 463
(1991).}%
\nref\rNiYo{Nishigaki S. and Yoneya T., Nucl. Phys. {\bf B348}, 787
(1991).}%
\nref\rVKO{Di Vecchia P., Kato M. and Ohta N., Nucl. Phys. {\bf B357}, 495
(1991).}
\nref\rDavidetal{David F., Nucl. Phys. {\bf
B257[FS14]}, 45, 543 (1985)\semi
Ambj{\o}rn J., Durhuus B. and Fr\"ohlich J., Nucl. Phys. {\bf B257[FS14]},
433 (1985)\semi Kazakov  V.A., Phys. Lett. {\bf B150}, 282
(1985).}\nref\rDS{Douglas M. and Shenker S., Nucl. Phys. {\bf B335}, 635
(1990)\semi Br\'ezin E. and Kazakov V., Phys. Lett. {\bf B236}, 144
(1990)\semi Gross D. and Migdal  A.A., Phys. Rev. Lett. {\bf 64}, 127
(1990); Nucl. Phys. {\bf
B340}, 333  (1990).}\nref\rD{Douglas M. R., Phys. Lett. {\bf B238}, 176
(1990)\semi David F., Mod. Phys. Lett. {\bf A5}, 1019  (1990)\semi Ginsparg
P., Goulian M., Plesser M.R. and Zinn-Justin J., Nucl. Phys. {\bf B342}, 539
(1990)\semi  Jevicki A. and Yoneya T.,  Mod. Phys. Lett. {\bf A5}, 1615
(1990).}
\nref\rAmbj{Ambj{\o}rn J.,  Durhuus  B. and Jonsson T., Phys. Lett. {\bf
B244}, 403  (1990).}
\newsec{Introduction}
Vector models have, in zero dimension, in the large $N$ limit, critical points
and scaling behaviours \refs{\rAMV,\rNiYo,\rVKO} that are reminiscent of
those of matrix models relevant to 2D quantum gravity \refs{\rDavidetal,
\rDS,\rD}. While matrix models are associated with surfaces because the
large $N$ limit selects planar Feynman diagrams, vector models are associated
with one-dimensional chains because only so-called ``bubble''
diagrams survive for  $N$ large. The corresponding critical points seem
indeed to have a natural physical
interpretation in the theory of branched polymers or filamentary surfaces
\refs{\rAmbj,\rAMV}. We report here the
extension of the analysis to  vector models in quantum mechanics and $d$
dimensional field theories \ref\rZinn{Zinn-Justin J., Phys. Lett. {\bf B257},
335  (1991); Saclay preprint  SPhT/91-054, Proceedings of the ICTP Workshop
on \lq\lq String Theory'' (Trieste 1991).}.\par
These peculiar critical points of vector models have the
following origin: In the large $N$ limit, vector models are reduced to
perturbation theory around minima of an effective potential resulting from the
competition   between an attractive potential and a measure term which
suppresses contributions of small fields. A singularity occurs when the
minimum of the effective potential disappears. In matrix models similarly
there is a competition between  the potential and a measure term which
produces a repulsion  between eigenvalues, although the detailed mechanism is
slightly different.\par
Vector and matrix models share also other properties:
Perturbation theory is always divergent \ref\rGZlob{Ginsparg P. and
Zinn-Justin J.,  Phys. Lett. {\bf B255}, 189  (1991)\semi Ginsparg P. and
Zinn-Justin J., ``Action principle and large order behaviour of
non-perturbative gravity'',
LA-UR-90-3687/SPhT/90-140 (1990), to appear in {\it Random surfaces and
quantum gravity\/}, proceedings of 1990 Carg\`ese workshop, edited by
O. Alvarez, E. Marinari, and P. Windey.} and
half of the vector models (like for matrix models) are ill-defined
beyond perturbation theory because perturbation theory is non-Borel summable
and the corresponding potential is not bounded from below. This is a serious
problem because in both cases this class includes the models with positive
weights, the only ones which have a simple physical interpretation.
\newsec{$O(N)$-invariant integrals}
{\it Branched polymers and $O(N)$-invariant models.} Consider the integral
$$Z=\int\d^{N} {\bf x}\,\d\sigma\,\exp\left(-{\bf x}^2-\sigma{\bf
x}^2-N\sigma^2/\civ\right).$$
We can integrate over the $N$-component vector $\bf x$:
$$Z=\int\d\sigma\exp\left[-N\left({\sigma^2/\lambda}+\ud\ln(1+\sigma)\right)
\right].$$
If we then expand the partition function in powers of $\civ$, we observe that
the  coefficients of the expansion are sums of Feynman diagrams which can be
interpreted as some kind of branched polymers. These polymers are weighted in
particular by a factor $\civ^l$ where $l$ counts the number of links
(generated by the
$\sigma$-propagator) and a  factor $N^{1-L}$ where $L$ is the number of
loops. As in matrix models the continuum limit is reached when $\civ$
approaches the singularity of $Z$ closest to the origin, emphasizing the
long chains. In the large $N$ limit only tree-like chains are selected. We
shall verify that there exists a double scaling limit in which chains with
an arbitrary number of loops contribute. \par
To see the connection with $O(N)$-invariant vector models we instead
integrate over $\sigma$ and obtain
$$Z=\int\d^{N} {\bf x}\,\exp\left[-{\bf x}^2+\civ \left({\bf x}^2\right)^2
/4 \right].$$
\medskip
{\it General $O(N)$-invariant vector models.} We immediately  consider
more general $O(N)$-invariant vector models with a partition function given by
the $N$-dimensional integral:
\eqn\eint{Z=\int\d^{N} {\bf x}\,\e^{-NV({\bf x}^2)},}
where  $V(\sigma)$ a polynomial. Integrating  over angular variables we
remain with an integral over $\sigma=x^2$
\eqn\eintb{Z\propto \int^{\infty}_0 {\d \sigma\over
\sigma}\e^{-N[V(\sigma)-{1\over 2}\ln \sigma]}.}
With this more general integrand we can vary independently the weights
associated with the different vertices. The large $N$ limit can be evaluated
by steepest descent. The saddle point $\sigma_s$ is given by:
\eqn\esadd{2V'\left(\sigma_s\right)\sigma_s=1\,. }
Critical points are points where the derivative of $V'(\sigma)\sigma$ at
$\sigma_s$ vanishes so that the result, at leading order, is no longer
given by the gaussian approximation. Let us assume more generally that $p-1$
successive derivatives vanish. The critical potential $V=V_c$ then
satisfies
$$2V'_c(\sigma)\sigma-1=O\bigl((\sigma_s-\sigma)^p\bigr),\ \Rightarrow\
V_c(\sigma)-\ud \ln\sigma={\rm const.}\
+O\bigl((\sigma_s-\sigma)^{p+1}\bigr).$$
We see that the values of $\sigma-\sigma_s$ contributing to the
integral are of order $N^{1/(p+1)}$. Let us add to  the  critical function
$V_c$ the set of relevant perturbations (in the sense of critical phenomena)
$$V(\sigma)=V_c(\sigma)+\sum_{q=1}^{p-1}v_q \left(\sigma_s-\sigma\right)^q \,
,$$
(the term $q=p$ can be eliminated by a shift of $\sigma$). Rescaling
$\sigma_s-\sigma$ into $(\sigma_s-\sigma) N^{-1/(p+1)}$ we see that the
scaling region  corresponds to take  $v_q=O( N^{(q-p-1)/(p+1)})$. Setting then
$$ \sigma_s-\sigma=zN^{-1/(p+1)},\qquad u_q=N^{(p+1-q)/(p+1)} v_q\,,$$
we find the {\it universal scaling} free energy $F=\ln Z$:
\eqn\elogZ{F(u_q)\sim \ln \left[\int \d z\,\exp\left(-
z^{p+1}+\sum_{q=1}^{p-1}u_q z^q \right)\right].}
Note that the case $p$ odd corresponds to convergent integrals while the
case $p$  even can be only reached by analytic continuation and yields
complex functions. This is in particular the case for the $p=2$ critical
point which is the only one which has a simple statistical interpretation
because all weights are positive. A similar situation is found in matrix
models, where in particular the unitary models are all unstable.
\newsec{$O(N)$ invariant potentials in quantum mechanics}
We consider now the partition function $Z$
\eqn\epartmq{Z=\int[\d {\bf x}(t)]\,\e^{-S(x)},}
for the $O(N)$ invariant action
\eqn\eactmq{S=N\int \d t \left[\ud \left(\dot{\bf x}\right)^2+\ud {\bf
x}^2+{\civ \over 4} \left( {\bf x}^2 \right)^2 \right].}
At the end of the section we shall briefly examine the problem of the
existence of a large $N$ scaling limit, using radial
coordinates, from the Schr\"odinger equation point of view (the only method
available in the matrix case). However, because we want to extend the
analysis to field theory, we use here a more general method, which directly
shows the relation between the vector model and a model of branched polymers
embedded in one dimension. We start from the identity
\eqn\eHub{\exp\left[-{N\civ \over 4}\int\d t\, \left( {\bf x}^{2} \right)^{2}
\right]
\propto \int\left[\d \sigma\right(t)] \exp\left[N\int\d t \left({\sigma^{2}
\over 4\civ}- {\sigma \over 2} {\bf x}^{2} \right)\right] .}
Introducing identity \eHub\ into the path integral,  we can integrate over
$\bf x$ and find
\eqn\eZeff{Z\propto \int [\d \sigma(t)]\,\e^{-S_{\rm eff}(\sigma)}, }
with
\eqn\eSeff{S_{\rm eff}(\sigma)={N \over 2}\left[-\int \d t
{\sigma^2(t)\over2\civ} +\tr \ln \left(-\d^2+1+\sigma\right)\right].}
The dependence of the partition function  on $N$ is now explicit and the
partition function can be interpreted in terms of a branched polymer
model, the $\sigma$ internal lines representing the polymer links.\par
In the large $N$ limit the path integral can be calculated by steepest
descent, $1/N$ playing the role of $\hbar$ in the classical limit. We
look for a saddle point $\sigma(t)=\sigma$ constant which is then a solution
of the equation:
\eqn\elamsadd{ {\sigma\over \civ} -{1 \over 2\pi}\int {\d \omega \over
\omega^2 +1 +\sigma} = {\sigma\over \civ}-{1 \over 2\sqrt{1+\sigma}}=0\,.}
A critical point is found when two different solutions meet. The condition
is
\eqn\econd{{\d \over \d\sigma}\left[{\sigma\over \civ}-{1 \over
2\sqrt{1+\sigma}}\right]={1\over \civ}+{1 \over 4 (1+ \sigma)^{3/2}}=0\, .}
This equation implies that $\civ$ is negative, and thus the potential is not
bounded from below. The analysis which follows has thus no meaning beyond
perturbation theory. Although the sum of the perturbative expansion can be
defined by analytic continuation, the result is then complex and thus
physically not acceptable.\par
With our normalizations the explicit values are
\eqn\ecivc{\civ_c=-4/3^{3/2}<0\, , \qquad \sigma_c=-2/3\,.}
\medskip
{\it The $1/N$-expansion.} To generate  the $1/N$ expansion we have first to
calculate the $\sigma$-propagator. It is convenient to set:
$$\mu^2=1+ \sigma \ \Rightarrow\ \mu^2_c=1/3.$$
The  Fourier transform $\Delta_{\sigma}(\omega)$ of the
$\sigma$-propagator is then:
\eqn\eprop{ \Delta^{-1}_{\sigma}(\omega)=-{N \over 2}\left[{1\over\civ }+{1
\over 2\pi}\int {\d \omega'\over
\left(\omega'{}^2+\mu^2\right)\left[\left(\omega-\omega' \right)^2
+\mu^2\right]} \right].}
The criticality condition implies that the inverse propagator vanishes at
$\omega=0$. Note that since $\mu_c$ does not vanish the inverse-propagator
remains a regular function of $\omega^2$, which behaves like $\omega^2$ for
$\omega$ small. In higher dimensions in the critical limit $\sigma$ would
become a massless field . In one dimension, however,
quantum fluctuations always lift the degeneracy of the classical ground state.
To understand the critical limit we perform a local (small $\omega$ in
Fourier transform) and small $\sigma-\sigma_c$ expansion of the action. We
then rescale time and $\sigma$:
\eqn\erescqm{t\mapsto t N^{1/5} \Longleftrightarrow \omega \mapsto
 \omega N^{-1/5}\,, \qquad \sigma-\sigma_c\mapsto\sigmt
N^{-2/5},}
to render the coefficients of $\sigmt^3$ and $(\dot{\sigmt})^2$ in the
action $N$ independent. We note that terms with higher derivatives
or higher powers of $\sigmt$ are then suppressed for $N$ large. Therefore at
leading order for $N$ large (and after some additional finite rescaling) the
action takes the form:
\eqn\eactiii{ S(\sigmt)\sim\int\d t \left[{1\over2}\left( {\d\sigmt\over \d
t}\right) ^2+{\sigmt^3\over 3}\right].}
The action is not bounded from below and the path integral can
be defined only by analytic continuation. The  corresponding the hamiltonian
$H$ is:
$$H=N^{-1/5}\left(-\ud (\d/\d\sigmt)^2+{\textstyle{1\over3}}\sigmt^3\right).$$
\medskip
{\it The scaling region.} For $\civ$ close to $\civ_c$
the most relevant new interaction term is the term linear in $\sigmt$.
After the rescaling \erescqm\ it gives an additional contribution to the
action proportional to $(\civ-\civ_c)N^{4/5}\sigmt$.  The scaling region is
thus defined by letting $N$ go to infinity and $\civ-\civ_c$ to zero at
$$(\civ-\civ_c)N^{4/5}=u $$
fixed. The hamiltonian relevant to the scaling limit is then:
\eqn\ehamiii{H=N^{-1/5}\left[-{1\over2} \left(\d\over\d\sigmt\right)^2+
u\sigmt+ {\sigmt^3\over3}\right].}
We have found a scaling regime analogous to the one observed in $d<1$ matrix
models. In the $d=1$ matrix model, instead, the situation is more complicated
\ref\roneD{Ginsparg  P. and Zinn-Justin J., Phys. Lett. {\bf B240}, 333
(1990)\semi Br\'ezin E., Kazakov  V. A., and Zamolodchikov Al. B.,
Nucl. Phys. {\bf B338}, 673  (1990)\semi
Parisi G., Phys. Lett. {\bf B238}, 209, 213  (1990);
Europhys. Lett. {\bf 11}, 595  (1990)\semi
Gross  D. J. and Miljkovic N., Phys. Lett. {\bf B238}, 217  (1990).} because
logarithmic deviations from a simple scaling law are found, a situation we
shall meet in the $d=2$ vector model.
\medskip
{\it Generalization.} The previous analysis can be extended to more general
$O(N)$-invariant potentials. Let us consider the action:
\eqn\eMQgen{S=N\int \d t \left[\ud \left(\dot{\bf x}\right)^2+V\left({\bf
x}^2\right)\right] .}
The large $N$ expansion is generated by standard techniques  (see for example
\ref\rbook{Zinn-Justin J., {\it Quantum
field theory and critical phenomena}, Oxford Univ. Press 1989.}).
We introduce a Lagrange multiplier $\rho$ to impose the constraint
$\sigma={\bf x}^2$. The action then takes the form
\eqn\eSlammu{ S({\bf x},\rho,\sigma)=N\int \d t \left[\ud \left(\dot{\bf
x}\right)^2 + V(\sigma) +\ud \rho\left({\bf x}^2-\sigma\right) \right].}
We integrate over $\bf x$ to obtain:
\eqn\eSefflm{ S(\rho,\sigma)=N\left\{\int \d t
\left[V(\sigma)-\ud\rho\sigma\right] +\ud
\tr\ln \left(-\d^2+ \rho\right)\right\}.}
In the large $N$ limit  the path integral can again be calculated by steepest
descent. We look for two constants $\rho,\sigma$ solutions of
\eqn\esaddlm{ V'(\sigma)-\ud \rho=0\, ,\qquad -\ud \sigma+{1 \over 4\pi}\int
{\d \omega \over \omega^2 +\rho}= -\ud \sigma+{1\over 4\sqrt{\rho}}=0\,.}
When the determinant of partial derivatives  of this system vanishes we find
a critical point:
$$\det S^{(2)}\equiv \det\pmatrix{V''(\sigma)&-\ud \cr -\ud &
-{\textstyle{1\over 8}}\rho^{-3/2} \cr}=0\,,$$
and thus
$$8\sigma^2 V'(\sigma)=1,\qquad 4\sigma^3 V''(\sigma)=-1.$$
The matrix  $S^{(2)}$ is the inverse propagator at $\omega=0$.
These conditions thus imply that the propagator has a pole  at $\omega=0$.
However only the linear combination $\sigmt$ of $\rho$ and $\sigma$ which
corresponds to the eigenvector of $S^{(2)}$ with zero eigenvalue is
singular. The non-zero mode can be integrated out and we find an effective
action for $\sigmt$. The main difference with
the case of the quartic interaction is that we can now adjust the original
potential $V$ in such a way that the coefficient of the
leading ${\sigmt}^3$  interaction vanishes, or more generally all
interactions up to the $\sigmt^{p}$ vanish. We then perform a local
expansion of the action and rescale time and
$\sigmt$ to render the  coefficients of $(\dot{\sigmt})^2$ and $\sigmt^{p+1}$
$N$ independent:
\eqn\erescp{t\mapsto t N^{(p-1)/(p+3)} \Longleftrightarrow \omega \mapsto
 \omega N^{-(p-1)/(p+3)}\,, \qquad \sigmt-\sigmt_c\mapsto\sigmt
N^{-2/(p+3)}.}
It is easy to verify that interactions containing derivatives or higher
powers of $\sigmt$ are then suppressed for $N$ large.\par
To describe the scaling region we can add to the critical potential a set of
relevant terms characterized by parameters $v_q$. The scaling limit is then
obtained by keeping the products $u_q$,
$$u_q=N^{(2p+2-2q)/(p+3)}v_q,\qquad q=1,\ldots,p-1\,,$$
fixed and the corresponding scaling hamiltonian is:
\eqn\ehamp{H=N^{-(p-1)/(p+3)}\left[-{1\over2} \left(\d\over\d\sigmt\right)^2+
{\sigmt^{p+1}\over p+1}+\sum_{q=1}^{p-1}u_q {\sigmt^q\over q} \right].}
Again only half of the models corresponding to $p$ odd are stable.
\medskip
{\it Hamiltonian formalism.} For $N$ large the zero angular momentum
hamiltonian $H_0$ can be written (after factorizing $r^{(N-1)/2}$ in the
wave function):
\eqn\eHzero{H_0=-{1\over2N}\left({\d \over \d r}\right)^2 +N W(r),}
where $r=|\bf x|$ and for $N$ large $W(r)$ is related to the potential
$V(r)$ by
$$W(r)={1 \over 8r^2}+V(r).$$
In the case of the anharmonic oscillator \eactmq\ for example
$$V(r)={r^2 \over 2}+\civ {r^4 \over4},$$
For $N$ large, the eigenvalues can be calculated by expanding perturbation
theory around the classical minimum $r_c$ of the potential $W$. A critical
potential is defined by the condition that the second derivative of $W$ also
vanishes:
\eqn\ecritco{W'(r_c)=W''(r_c)=0\,.}
If moreover $p$ derivatives of $W$ vanish the leading order hamiltonian is
$$H_{\rm c}=NW(r_c)-{1\over2N}\left({\d \over \d r}\right)^2 +{N\over
(p+1)!}W^{(p+1)}(r_c)(r-r_c)^{p+1},$$
which leads after rescaling to a critical contribution to the ground state
energy of order $ N^{-(p-1)/(p-3)}$ in agreement with \ehamp.\par
In the special case \eactmq\ conditions \ecritco\ yield $r^2=3^{1/2}/2$,
$\civ=-4/3^{3/2}$. We recognize the critical value \ecivc\ of $\civ$.
\newsec{Field theory: Critical points in the large $N$ limit}
We come now to the most interesting case, field theory in $d>1$ dimensions,
which we expect, according to the previous analysis, to correspond to
branched polymers embedded in  $d$-dimensional space.\par
In contrast to quantum mechanics ($d=1$) we expect now the $\sigma$-field
to remain massless even after taking into account the successive corrections
of the large $N$ expansion. The field $\sigma$, which is equivalent to
the composite $\phi^2 $ field, is associated with a massless bound state of
the field $\phi$, which itself remains non-critical.\par
We discuss the problem only in the special case of the
$(\phi^2)^2$ interaction, the extension to more general cases being
straightforward. We consider the partition function:
\eqn\eZfld{ Z = \int \left[ \d \phi(x) \right] \e^{-S ( \phi)},}
where $S(\phi)$ is the $ O(N) $ symmetric action:
\eqn\eactON{S( \phi)= N\int \left\{ { 1 \over 2} \left[\partial_{\mu} \phi (x)
\right]^{2}+{1 \over 2}r \phi^{2}(x)+{\civ \over 4} \left[ \phi^{2}
(x) \right]^{2} \right\} \d^{d}x. }
A cut-off of order $1$, consistent with the symmetry, is now implied. For
example  we can assume that the inverse propagator has higher order
derivative terms and we have explicitly  written in action \eactON\ only the
two first terms in a local expansion (in  Fourier space small
momentum expansion). In particular, the parameter $r$ is then the
value of the inverse propagator at zero momentum. \par
We use the same algebraic identity as in the quantum mechanical case
\eqn\eHubi{ \exp\left[-{N\civ \over 4}\int\d^{d} x\, \left( \phi^{2}
\right)^{2}\right]
\propto \int\left[\d\sigma(x)\right] \exp\left[N\int\d^{d} x \left({\sigma^{2}
\over 4\civ}-{\sigma \over 2} {\bf \phi}^{2} \right)\right] .  }
Introducing this identity into \eZfld\ and integrating over $\phi $ we find
\eqn\eZfldb{ Z= \int \left[ \d  \sigma(x)\right]  \e^{-S_{ {\rm
eff}}(\sigma)}, } with:
\eqn\eactef{S _{ \rm eff} (\sigma) ={N \over 2} \left[ -\int
{\sigma^{2}(x) \over 2\civ}
  \d^{d}x +\tr\ln \bigl( -\Delta +r+\sigma (x) \bigr) \right] , }
expression which shows that the vector model is now related to a branched
polymer problem in $d$ dimensions.\par
\medskip
{\it The large $ N $ limit.} In the large $N$ limit the functional integral
can be calculated by steepest descent. We look for a uniform saddle point
$\sigma(x)=\sigma$. The saddle point equation is:
\eqn\esadfld{ -{\sigma\over \civ} + {1 \over (2\pi)^{d}} \int{ \d ^{d}p \over
p^{2}+r+ \sigma }  = 0\, .  }
Let us introduce the parameter $ m $: $m^{2}=r+\sigma$, which is, in the
large $N$ limit, the mass of field $\phi$. Eq.\ \esadfld\ can then be
written:
\eqn\esadfb{ { \left(m^{2}-r \right)\over \civ} -{ 1\over \left(2\pi \right)
^{d}} \int{ \d ^{d}p \over p^{2}+m^{2}}  = 0\,. }
The solution is singular when the derivative with respect to $\sigma$ or
$m^2$ vanishes. This yields the equation for the critical point:
\eqn\ecritft{{1\over \civ}+ {1 \over \left(2\pi \right)
^{d}} \int{ \d ^{d}p \over\left(p^{2}+m^{2} \right)^2}=0\, . }
Eqs.\ \esadfld,\ecritft\ define, at $r$ fixed, critical values of $\civ$
and $\sigma$. The critical value  $\civ_c$ of $\civ$ is again negative.
Criticality eq.\ \ecritft\ implies that the $\sigma$-propagator has a
pole at zero momentum. The $\sigma$-field becomes massless while the
$\phi$-field remains massive.
\newsec{The double scaling limit}
We have studied the large $N$ limit. We now look for a scaling limit.
Since the $\sigma$ field is a one-component field it can remain critical
for $d>1$ even in presence of interactions. Still perturbation theory is IR
divergent for dimensions $d\le6$. We therefore add a relevant
perturbation proportional to $\civ-\civ_c$ which provides the theory with
an IR cut-off. We then look for the most IR divergent terms in
perturbation theory: This is a  problem standard in the
theory of critical phenomena \refs{\rbook}, the bare mass squared which is
proportional to $ \sqrt{\civ_c-\civ}$  playing the role of a
deviation from the critical temperature. The effective action for the
$\sigma$-field is non-local and contains arbitrary
powers of the field. However, because the $\phi$-field is not
critical  we can again make a local expansion. Standard arguments of the
theory of critical phenomena tell us that the most IR divergent terms come
from interactions without derivatives and with the lowest power of the field.
Here the leading interaction is proportional to $\sigma^3$.
To characterize the IR divergences of the perturbative expansion in powers in
$1/N$ we again rescale distances and field $\sigma-\sigma_c$:
\eqn\erescali{ \sigma-\sigma_c \propto \sigmt
\Lambda^{(2-d)/2}N^{-1/2},\qquad x\mapsto \Lambda x,}
where $\Lambda$ will play the role of a cut-off. The effective action at
leading order, after some additional finite renormalizations, is
$$S_{\rm eff}(\sigmt)=\int\d^d x\left[\ud
\left(\partial_{\mu}\sigmt\right)^2  + v \Lambda^{(d+2)/2} \sqrt{N}
\sigmt + {\Lambda^{(6-d)/2} \over 3\sqrt{N}}\sigmt^3
\right],$$
where $v\propto \civ-\civ_c$ and a cut-off $\Lambda$ is now implied. \par
We  first consider dimensions $d<6$. The $\sigmt^3$ field theory is then
super-renormalizable and we fix the coefficient of $\sigmt^3$:
$$\Lambda^{(6-d)/2}/\sqrt{N}= g_3.$$
Therefore the cut-off grows with $N$ like $N^{1/(6-d)}$.
However, unlike the case of quantum mechanics, we cannot also  keep the
quantity $ v \Lambda^{(d+2)/2}  \sqrt{N} $ fixed because the field theory has
UV divergences when the cut-off $\Lambda$ becomes large.
\medskip
{\it Dimensions $d<4$.}
It is convenient to examine first dimensions $d<4$ because then only the field
average is divergent. We have to introduce a counterterm which renders
$<\sigmt(x)>$ finite. The renormalized action is
$$S_{\rm eff}(\sigmt_r)=\int\d^d x\left[\ud
\left(\partial_{\mu}\sigmt_r\right)^2  +\ud \mu^2\sigmt_r^2 +
{g_3\over3}\sigmt_r^3 -c_1(\Lambda)\sigmt_r\right],$$
in which $\sigmt_r$ is the renormalized field and $\mu$ is a renormalized
mass parameter.  \par
To recover the original action we must eliminate the term quadratic in the
field and thus shift $\sigmt_r$ by a quantity $\bar\sigma$:
$$\bar\sigma=-\mu^2/2g_3.$$
Identifying then the coefficients of the linear term we find:
$$ v \Lambda^{(d+2)/2} \sqrt{N}= -c_1(\Lambda)-{\mu^4 \over 4g_3}.$$
For $d=2$ only the one-loop diagram is divergent and we obtain
$$c_1(\Lambda)={g_3\over2\pi}\ln(\Lambda/\mu).$$
Therefore to obtain a non-trivial scaling limit we have to
choose:
\eqn\ereltwo{v=-{1\over N}\left[{1\over8\pi}\ln(Ng_3^2/\mu^4)+{\mu^4\over
4g_3^2}\right].}
Note that only the $\ln N/N$ term is universal, the $1/N$ term is
regularization and renormalization dependent.\par
For  $d=3$  the two-loop diagram is also divergent. Still
for $\Lambda$ large the leading contribution is still given by the one-loop
diagram, thus  $c_1(\Lambda)\propto \Lambda$ and $v=O(1/N)$.
We have thus shown that for $d<4$ a scaling limit exists which leads to a
renormalized $\sigma^3$ field theory. The relation between $\civ-\civ_c$ and
$N$, however, has itself no longer a simple power law form.\par
Note finally the similarity between the results for the vector model at $d=2$
and the matrix model at $d=1$ \refs{\roneD}.
\medskip
{\it Higher dimensions.} For $4\le d<6$ the situation is slightly more
complicated because two counterterms are required, renormalizing $<\sigmt>$
and the coefficient of $\sigmt^2$. The quantity $v\propto (\civ-\civ_c)$
becomes a even more complicated function of $N$. However, at leading order
for $N$ large, $v$ still behaves as $1/N$, while naive scaling would
have predicted a scaling variable $N v^{(6-d)/4}$. \par
For $d\ge 6$ IR divergences are no longer strong enough to compensate the
$1/N$ factors and thus no non-trivial scaling limit can be defined.
\medskip
{\it More General Interactions.}
The method applicable to more general interactions has already been
explained in the case of quantum mechanics. Because only one mode is critical
the main effect is to introduce additional parameters in the effective
interaction of the critical field in such a way that the most IR divergent
interactions can be
cancelled. In the language of critical phenomena we reach multicritical
points. In dimensions $d<2(p+1)/(p-1)$ we then generate the renormalized
$\sigma^{p+1}$ interactions provided we again choose the parameters  of the
potential, as functions of $N$, such that they cancel the UV divergences of
perturbation theory.
\newsec{Conclusion, prospects}
We have shown on a few simple examples that vector models have, at least in
low dimensions, non-trivial large $N$ scaling limits. Obviously more general
models can be analyzed by similar methods, for example  by introducing
several vector fields additional degrees of freedom for the polymers are
generated. Of particular interest is the generalization  to models containing
fermions (in particular supersymmetric
models). One would like to find out whether the presence of fermions
stabilizes some of the models which are unstable otherwise, a very serious
problem  for the corresponding matrix models of 2D gravity.
\listrefs
\bye